\documentclass[11pt]{article}
\newcommand{\be}{\begin{equation}}
\newcommand{\ee}{\end{equation}}
\newcommand{\bea}{\begin{eqnarray}}
\newcommand{\eea}{\end{eqnarray}}
\newcommand{\sn}{{\rm sn}}

\newcommand{\dn}{{\rm dn}}
\newcommand{\cn}{{\rm cn}}
\newcommand{\sech}{{\rm sech}}

\begin{document}

\vspace{0.5in}
\begin{center}
{\LARGE{\bf Novel Superposed Kink and Pulse Solutions for $\phi^4$, MKdV,
NLS and Other Nonlinear Equations}} 
\end{center}

\begin{center}
{\LARGE{\bf Avinash Khare}} \\
{Physics Department, Savitribai Phule Pune University \\
 Pune 411007, India}
\end{center}

\begin{center}
{\LARGE{\bf Avadh Saxena}} \\ 
{Theoretical Division and Center for Nonlinear Studies, 
Los Alamos National Laboratory, Los Alamos, New Mexico 87545, USA}
\end{center}

\vspace{0.9in}
\noindent{\bf {Abstract:}}

We show that a number of nonlinear equations including symmetric as well as 
asymmetric $\phi^4$, modified Korteweg de Vries (MKdV), mixed KdV-MKdV, 
nonlinear Schr\"odinger (NLS), quadratic-cubic NLS as well as higher order 
neutral scalar field theories, higher order KdV-MKdV and higher order 
quadratic-cubic NLS admit superposed periodic kink and pulse solutions and 
some of them also admit superposed hyperbolic kink solutions. 

\section{Introduction}

During the last few decades several nonlinear equations including $\phi^4$, 
NLS, KdV and MKdV equations \cite{dj} have found application in several 
different areas of physics. While some of them are integrable, the others are 
nonintegrable. All these systems admit periodic as well as hyperbolic
pulse and kink solutions. Besides, the integrable models such as NLS, KdV, 
MKdV and some others also admit N-soliton solutions, breather solutions, etc. 
Many novel properties of the nonlinear equations have been discovered over 
the years. However, the nonlinear equations are very rich and it is not clear 
whether we have uncovered all of their novel properties.

Several years ago, in a largely unnoticed paper, Tankeyev, Smagin, 
Borich and Zhuravlev \cite{tan} obtained a novel superposed hyperbolic kink 
solution for the repulsive MKdV equation. Their work was inspired by the
earlier work of \cite{per}. The obvious question is whether a similar solution 
also exists for other nonlinear equations. Secondly, are there superposed
periodic as well as hyperbolic pulse and kink solutions for MKdV and other 
nonlinear equations? The purpose of this paper is to answer these questions  
in the affirmative. In particular, we show that a large number of 
nonlinear equations, including both symmetric and asymmetric $\phi^4$, 
MKdV, NLS and several other nonlinear equations, many of which
have found wide application, indeed admit superposed periodic and hyperbolic 
kink and periodic pulse solutions. By {\it superposed} here we mean that a 
solution is a linear combination of two kink solutions or two pulse solutions or 
the corresponding periodic solutions. Interestingly, such solutions are known 
to exist in both condensed matter and field theory contexts, e.g. the scalar 
potential for baryons or polarons in conducting polymers \cite{dashen, 
campbell, saxena, thies}. 

The plan of the paper is the following. In Sec. II we show that the 
celebrated symmetric double well 
$\phi^4$ equation admits superposed periodic kink and pulse solutions. 
In Sec. III we consider the asymmetric double well $\phi^4$ equation and show 
that it not only admits the 
superposed periodic kink and pulse solutions but also admits two distinct 
types of hyperbolic 
superposed kink solutions. We also consider a one-parameter family of 
scalar field theories
of the form $\phi^{2}$-$\phi^{2m+2}$-$\phi^{4m+2}$ and show that even these
models admit superposed hyperbolic kink solutions. In Sec. IV we discuss 
the MKdV equation and show that apart from the superposed hyperbolic kink solution 
already known \cite{tan}, like the symmetric $\phi^4$ case, the MKdV equation also 
admits superposed periodic kink and pulse solutions. Further, using the celebrated 
Miura transformation \cite{miura}, from the superposed solutions of the 
repulsive MKdV equation we immediately obtain the corresponding 
superposed solutions of the KdV equation. In Sec. V we discuss mixed 
KdV-MKdV equation \cite{mixed} 
and show that like the asymmetric $\phi^4$ case, it also admits superposed 
periodic kink and pulse
solutions as well as superposed hyperbolic kink solutions. We further 
consider a one-parameter family of generalized mixed KdV-MKdV system 
characterized by 
$u^{m} u_{x}$-$u^{2m} u_{x}$ ($m = 1,2,3,...$) and show that it admits 
novel superposed hyperbolic kink solutions.
In Sec. VI we consider the celebrated integrable NLS equation
and show that like the symmetric $\phi^4$ case, it also admits novel 
superposed periodic kink 
and pulse solutions. In Sec. VII we study a mixed quadratic-cubic NLS 
equation \cite{quad} and show that like the asymmetric $\phi^4$ case, it not 
only admits superposed periodic pulse and kink solutions but also admits  
a superposed hyperbolic kink solution. In addition, we consider a one-parameter family  
of generalized higher order NLS systems characterized by $|u|^{m} u$-$|u|^{2m} u$ 
($m = 1,2,3,...$) and show that even such
systems admit superposed hyperbolic kink solutions. Finally, in Sec. VIII 
we summarize
the main results obtained in this paper and discuss some of the open problems.

\section{Superposed Solutions of Symmetric $\phi^4$ Field Theory}

In this section we show that the well known $\phi^4$ field equation
\be\label{1.1}
\phi_{xx} = a \phi + d \phi^3\,,
\ee
not only admits the periodic kink and the pulse solutions $\sn(x,m)$ 
$\cn(x,m)$ and $\dn(x,m)$
respectively, but even superposed periodic kink and pulse solutions. 
Here $\sn(x,m)$, $\cn(x,m)$ and $\dn(x,m)$ denote Jacobi elliptic 
functions with modulus $m$ \cite{as}. 
 
But before that let us 
recall the three well known periodic solutions of the $\phi^2$-$\phi^4$ 
field Eq. (\ref{1.1}).
In particular, it is well known \cite{aub, ks1} that Eq. (\ref{1.1}) admits 
the pulse-like periodic solutions 
\be\label{1.2}
\phi(x) = A \cn(\beta x,m)\,,
\ee
provided
\be\label{1.3}
d A^2 = -2m \beta^2\,,~~a = (2m-1)\beta^2\,,
\ee
and 
\be\label{1.4}
\phi(x) = A \dn(\beta x,m)\,,
\ee
provided
\be\label{1.5}
d A^2 = -2\beta^2\,,~~a = (2-m)\beta^2\,.
\ee
In the limit $m = 1$, both the solutions (\ref{1.2}) and (\ref{1.4}) go over 
to the hyperbolic pulse solution
\be\label{1.6}
\phi(x) = A \sech(\beta x)\,,
\ee
provided
\be\label{1.7}
d A^2 = -2\beta^2\,,~~a = \beta^2\,.
\ee
Notice that for all three pulse solutions, $a > 0, d < 0$.

The same field Eq. (\ref{1.1}) also admits the periodic kink solution
\be\label{1.8}
\phi(x) = A  \sn(\beta x,m)\,,
\ee
provided
\be\label{1.9}
d A^2 = 2 m\beta^2\,,~~a = -(1+m)\beta^2\,. 
\ee
In the limit $m = 1$, the periodic kink  solution (\ref{1.8}) goes over to the
celebrated kink solution
\be\label{1.10}
\phi(x) = A \tanh(\beta x)\,,
\ee
provided
\be\label{1.11}
d A^2 = 2\beta^2\,,~~a = -2\beta^2\,.
\ee
Notice that unlike the three pulse solutions, the two kink solutions are valid
if $a < 0, d > 0$.

We now show that the symmetric $\phi^4$ field Eq. (\ref{1.1}) also admits four 
novel  periodic solutions which can be written as superposition of either the
periodic kink or the pulse solutions $\sn(x,m)$ and $\dn(x,m)$, respectively.

{\bf Solution I}

It is easy to check that the $\phi^4$ field Eq. (\ref{1.1})  admits 
the periodic solution
\be\label{1.12}
\phi(x) = \frac{A \dn(\beta x,m) \cn(\beta x,m)}{1+B\cn^2(\beta x,m)}\,,
~~B > 0\,,
\ee
provided
\bea\label{1.13}
&&0 < m < 1\,,~~a = [5m-1 -6B(1-m)]\beta^2\,, \nonumber \\
&&d m A^2 = [2m B(1-2B)+6(1-m) B^2]\beta^2\,,
\eea 
while $B$ satisfies a cubic equation
\be\label{1.14}
(1-m)^2 B^3 -3m(1-m)B^2 +B m (3m-1) +m^2 = 0\,.
\ee
Note that this solution is not valid for $m = 1$, i.e. the symmetric
$\phi^4$ field Eq. (\ref{1.1}) does not admit a corresponding hyperbolic 
solution.

This cubic equation is easily solved once one realizes that one solution must
be $B = m/(1-m)$ giving the well known solution 
$\phi = \frac{A \cn(\beta x,m)}{\dn(\beta x,m)}$. 
We find that the three roots of the cubic equation are
\be\label{1.15}
B = \frac{m}{1-m}\,,~~\pm \frac{\sqrt{m}}{1-\sqrt{m}}\,,
\ee
out of which the acceptable solution is
\be\label{1.16}
B = \frac{\sqrt{m}}{1-\sqrt{m}}\,,
\ee
and the corresponding values of $a$ and $d A^2$ turn out to be
\be\label{1.17}
a = -[1+m+6\sqrt{m}]\beta^2 < 0\,,~~
d A^2 = \frac{8 \sqrt{m} \beta^2}{(1-\sqrt{m})^2}\,.
\ee
Notice that for this solution $a < 0, d > 0$. 

At this stage, we recall the well known addition theorem for $\sn(x,m)$ 
\cite{as}, i.e.
\be\label{1.18}
\sn(a+b,m) = \frac{\sn(a,m) \cn(b,m) \dn(b,m) + \cn(a,m) \dn(a,m) \sn(b,m)}
{1-m \sn^2(a,m) \sn^2(b,m)}\,.
\ee
From here it is straightforward to obtain the identity
\be\label{1.19}
\sn(y+\Delta,m)-\sn(y-\Delta,m) = \frac{2\cn(y,m) \dn(y,m) \frac{\sn(\Delta,m)}
{\dn^2(\Delta,m)}}{1+B\cn^2(y,m)}\,,~~B = \frac{m \sn^2(\Delta,m)}
{\dn^2(\Delta,m)}\,.
\ee
On comparing Eqs. (\ref{1.12}) and (\ref{1.19}) and using Eqs. (\ref{1.16}) and
(\ref{1.17}), one can
re-express the periodic solution I given by Eq. (\ref{1.12}) 
as superposition of two periodic kink solutions, i.e.
\be\label{1.20}
\phi(x) = \frac{\sqrt{2m} \beta}{\sqrt{d}} 
\bigg [\sn(\beta x +\Delta, m) -\sn(\beta x- \Delta, m) \bigg ]\,,
\ee
Here $\Delta$ is defined by $\sn(\sqrt{m}\Delta,1/m) = \pm m^{1/4}$,
where use has been made of the identity
\be\label{1.21}
\sqrt{m} \sn(y,m) = \sn(\sqrt{m} y,1/m)\,. 
\ee

{\bf Solution II}

Remarkably, the symmetric $\phi^4$ Eq. (\ref{1.1}) also admits another periodic 
solution
\be\label{1.22}
\phi(x) = \frac{A \sn(\beta x,m)}{1+B\cn^2(\beta x,m)}\,,~~B > 0\,,
\ee
provided
\bea\label{1.23}
&&0 < m < 1\,,~~d A^2 = [-6B^2(1-m)+20 m B -8 B +14m]\beta^2\,,
\nonumber \\
&&a = [(5m-1) + \frac{6m}{B}] \beta^2\,,
\eea 
where $B$ satisfies a cubic equation
\be\label{1.24}
(1-m) B^3 -(3m-1)B^2-3 m B -m = 0\,.
\ee
Note that this solution is also not valid for $m = 1$, i.e. the symmetric
$\phi^4$ field Eq. (\ref{1.1}) does not admit a corresponding hyperbolic 
solution. This cubic equation is easily solved once one realizes that one 
root must be $B = -1$ giving rise to the well known singular
solution $\phi = A/\sn(\beta x,m)$. We find that the three roots of 
Eq. (\ref{1.24}) are
\be\label{1.25}
B = -1, \pm \frac{\sqrt{m}}{1-\sqrt{m}}\,,
\ee
out of which the only acceptable solution is 
\be\label{1.26}
B = \frac{\sqrt{m}}{1-\sqrt{m}}\,.
\ee
Using this value of $B$ one finds that
\be\label{1.27}
a = [6\sqrt{m}-(1+m)]\beta^2\,,~~d A^2 = -8\sqrt{m} \beta^2\,.
\ee
Thus for this solution while $d < 0$, $a > (<)$ 0 depending on if 
$6\sqrt{m} > (<)$ 0. 

Now on using the $\sn(x,m)$ addition theorem (\ref{1.18}), one can 
derive another novel identity
\be\label{1.28}
\sn(y+\Delta,m)+\sn(y-\Delta,m) = \frac{2\sn(y,m) \frac{\cn(\Delta,m)}
{\dn(\Delta,m)}}{1+B\cn^2(y,m)}\,,~~B = \frac{m \sn^2(\Delta,m)}
{\dn^2(\Delta,m)}\,.
\ee
On comparing Eqs. (\ref{1.22}) and (\ref{1.28}) and using Eqs. (\ref{1.26})
and (\ref{1.27}), the periodic solution II given by Eq. (\ref{1.22}) can be
re-expressed as superposition of two periodic kink solutions
\be\label{1.29}
\phi(x) = i\sqrt{\frac{2m}{|d|}}\beta  
\bigg [\sn(\beta x +\Delta, m)+\sn(\beta x -\Delta, m) \bigg ]\,,
\ee
Here $\Delta$ is defined by $\sn(\sqrt{m}\Delta,1/m) = \pm m^{1/4}$,
where use has been made of the identity (\ref{1.21}).

It is worth noting that for both the solutions I and II, the value of $B$ 
is the same but
while $d > 0$ for the first solution, $d <0$ for the second solution.

{\bf Solution III}

It is easy to check that the symmetric $\phi^4$ field Eq. (\ref{1.1}) 
admits another periodic solution
\be\label{1.30}
\phi(x) = \frac{A \sn(\beta x,m) \cn(\beta x,m)}{1+B\cn^2(\beta x,m)}\,,
\ee
provided
\bea\label{1.31}
&&0 < m < 1\,,~~d A^2 = [14(1-m)B^2 +12 B-20 B m -6m]\beta^2\,,~~ 
\nonumber \\
&&a = [5m-4-6(1-m)B] \beta^2\,,
\eea 
while $B$ satisfies a cubic equation
\be\label{1.32}
(1-m)B^3 +3(1-m)B^2 -(3m-2)B -m = 0\,.
\ee
Note that this solution is also not valid for $m = 1$, i.e. the symmetric
$\phi^4$ field Eq. (\ref{1.1}) does not admit a corresponding hyperbolic 
solution. The cubic Eq. (\ref{1.32}) is easily solved once one realizes 
that one solution must be $B = -1$ giving rise to the singular solution
$\phi = A \cn(\beta x,m)/\sn(\beta x,m)$. It turns out that the three roots of
$B$ are
\be\label{1.33}
B = -1\,,~~B = \frac{1-\sqrt{1-m}}{\sqrt{1-m}}\,,~~
B= -\frac{1+\sqrt{1-m}}{\sqrt{1-m}}\,,
\ee
out of which the only acceptable solution is
\be\label{1.34}
B = \frac{1-\sqrt{1-m}}{\sqrt{1-m}}\,.
\ee
Using this value of $B$ in Eq. (\ref{1.31}) we find that
\be\label{1.35}
a = (2-m-6\sqrt{1-m})\beta^2\,,~~ d A^2 
= -\frac{8(1-\sqrt{1-m})^2 \beta^2}{\sqrt{1-m}}\,.
\ee
Notice that for this solution $d < 0$ while $a$ could be positive or negative 
depending on the value of $m$. 

On using the well known addition theorem for $\dn(x,m)$ \cite{as}
\be\label{1.36}
\dn(a+b,m) = \frac{\dn(a,m) \dn(b,m) - m \sn(a,m) \cn(a,m) \sn(b,m) \cn(b,m)}
{1-m\sn^2(a,m) \sn^2(b,m)}\,,
\ee
one can derive the novel identity
\bea\label{1.37}
&&\dn(y-\Delta,m) -\dn(y+\Delta,m) = \frac{2m\sn(\Delta,m) \cn(\Delta,m) 
\sn(y,m) \cn(y,m)} {\dn^2(\Delta,m)[1+ B \cn^2(y)]}\,, \nonumber \\
&&B = \frac{m \sn^2(\Delta,m)}{\dn^2(\Delta,m)}\,. 
\eea

On comparing solutions (\ref{1.30}) and (\ref{1.37}) and using 
Eqs. (\ref{1.34}) and (\ref{1.35}), we find that the solution 
III of the symmetric $\phi^4$ field Eq. (\ref{1.1}) given by Eq. (\ref{1.30}) 
can be re-expressed as a 
superposition of two periodic pulse solutions, i.e.
\be\label{1.38}
\phi(x) = \beta \sqrt{\frac{2}{|d|}} \bigg (\dn[\beta x -\frac{K(m)}{2},m] 
- \dn[\beta x +\frac{K(m)}{2},m] \bigg )\,.
\ee

{\bf Solution IV}

Remarkably, the symmetric $\phi^4$ Eq. (\ref{1.1}) also admits another periodic 
solution
\be\label{1.39}
\phi(x) = \frac{A \dn(\beta x,m)}{1+B\cn^2(\beta x,m)}\,,
\ee
provided
\bea\label{1.40}
&&0 < m < 1\,,~~ m d A^2 = [6 B^2 (1-m)-20 m B +12 B -14m]\beta^2\,,
\nonumber \\
&&a = [(5m-4) + 6m/B] \beta^2\,,
\eea 
while $B$ satisfies a cubic equation
\be\label{1.41}
(1-m)^2 B^3 +B^2(1-m) (2-3m)-3 m(1-m) B + m^2 = 0\,.
\ee
Note that this solution is also not valid for $m = 1$, i.e. the symmetric
$\phi^4$ field Eq. (\ref{1.1}) does not admit a corresponding hyperbolic 
solution. This cubic equation is easily solved once one realizes that one 
root must be $B = \frac{m}{1-m}$. We find that the three roots of 
Eq. (\ref{1.41}) are
\be\label{1.42}
B = \frac{m}{1-m},~~ \pm \frac{1-\sqrt{1-m}}{\sqrt{1-m}}\,,
\ee
out of which the only acceptable solution is 
\be\label{1.43}
B =  \frac{1-\sqrt{1-m}}{\sqrt{1-m}}\,. 
\ee
Note that the choice $B = \frac{m}{1-m}$ leads to the well known solution 
$\phi = \frac{(1-m)A}{\dn(\beta x, m)}$.
Using $B$ as given by Eq. (\ref{1.43}), one finds that
\be\label{1.44}
a = [2-m+6\sqrt{1-m}]\beta^2\,,~~d A^2 = -\frac{8}{\sqrt{1-m}} \beta^2\,.
\ee
Thus for this solution while $d < 0$, $a > 0$.

Now using the addition theorem for $\dn(x,m)$ as given by Eq. (\ref{1.36}), 
one can derive another novel identity, i.e.
\be\label{1.45}
\dn(y+\Delta,m)+\dn(y-\Delta,m) = \frac{2\dn(y,m)}
{\dn(\Delta,m) [1+B\cn^2(y,m)]}\,,~~B = \frac{m \sn^2(\Delta,m)}
{\dn^2(\Delta,m)}\,.
\ee
On comparing Eqs. (\ref{1.39}) and (\ref{1.45}) and using Eqs. (\ref{1.43})
and (\ref{1.44}), the periodic solution IV given by Eq. (\ref{1.39}) can be
re-expressed as superposition of two periodic pulse solutions, i.e.
\be\label{1.46}
\phi(x) = \sqrt{\frac{2}{|d|}} \beta  
\bigg (\dn[\beta x +K(m)/2, m]+\dn[\beta x -K(m)/2, m] \bigg )\,.
\ee

It is worth noting that for both the superposed periodic pulse 
solutions III and IV, not only the 
value of $B$ is the same but also $d < 0$ for both the solutions. 
It is worth reminding that even for the periodic pulse solution
$\phi = A \dn(\beta x, m), d < 0$ (see Eqs. (\ref{1.4}) and (\ref{1.5})). 

\section{Superposed Solutions of Asymmetric $\phi^4$ Field Theory}

We now show that the asymmetric $\phi^4$ field equation
\be\label{2.1}
\phi_{xx} = a \phi - b \phi^2 + d\phi^3\,,
\ee
not only admits the superposed periodic kink and pulse solutions 
but even the superposed hyperbolic  kink solution. 
 
But before we proceed further, let us 
recall the three well known periodic solutions of the asymmetric 
field Eq. (\ref{2.1}) \cite{sanati, ks2}. In particular, it is well known that 
Eq. (\ref{2.1}) admits the periodic pulse solutions
\be\label{2.2}
\phi(x) = A+ B\cn(\beta x,m)\,,
\ee
provided
\be\label{2.3}
b = 3 dA\,,~~a = 2 dA^2\,,~~ d B^2 - 2m \beta^2\,,
~~d A^2 = -(2m-1)\beta^2\,,
\ee
which implies that this solution exists only if
\bea\label{2.4}
&&a, b, d < 0\,,~~2b^2 = 9 a d\,,~~m > 1/2\,,~~
A = \sqrt{\frac{|a|}{2|d|}}\,, \nonumber \\
&&B = \sqrt{\frac{|a|m}{(2m-1)|d|}}\,,~~
\beta = \sqrt{\frac{|a|}{2(2m-1)}}\,.
\eea

Similarly, Eq. (\ref{2.1}) also admits another periodic pulse solution
\be\label{2.5}
\phi(x) = A+ B\dn(\beta x,m)\,,
\ee
provided
\be\label{2.6}
b = 3 dA\,,~~a = 2 dA^2\,,~~ d B^2 = - 2 \beta^2\,,
~~d A^2 = -(2-m)\beta^2\,,
\ee
which implies that this solution exists only if
\bea\label{2.7}
&&a, b, d < 0\,,~~2b^2 = 9 a d\,,~~
A = \sqrt{\frac{|a|}{2|d|}}\,, \nonumber \\
&&B = \sqrt{\frac{|a|}{(2-m)|d|}}\,,~~
\beta = \sqrt{\frac{|a|}{2(2-m)}}\,.
\eea

In the limit $m = 1$ both solutions (\ref{2.2}) and (\ref{2.5}) go over to the
hyperbolic pulse solution
\be\label{2.8}
\phi(x) = A +B\sech(\beta x)\,,
\ee
provided
\be\label{2.9}
b = 3 dA\,,~~a = 2 dA^2\,,~~ d B^2 = -2 \beta^2\,,
~~d A^2 = -\beta^2\,,
\ee
which implies that this solution exists only if
\bea\label{2.10}
&&a, b, d < 0\,,~~2b^2 = 9 a d\,,~~
A = \sqrt{\frac{|a|}{2|d|}}\,, \nonumber \\
&&B = \sqrt{\frac{|a|}{|d|}}\,,~~\beta = \sqrt{\frac{|a|}{2}}\,.
\eea
Notice that for all the three pulse solutions, $a, b, d < 0$.

The same field Eq. (\ref{2.1}) also admits the periodic kink solution
\be\label{2.11}
\phi(x) = A +B \sn(\beta x,m)\,,
\ee
provided
\be\label{2.12}
b = 3 dA\,,~~a = 2 dA^2\,,~~ d B^2 =  2m \beta^2\,,
~~d A^2 = (1+m) \beta^2\,,
\ee
which implies that this solution exists only if
\bea\label{2.13}
&&a, b, d > 0\,,~~2b^2 = 9 a d\,,~~
A = \sqrt{\frac{a}{2d}}\,, \nonumber \\
&&B = \sqrt{\frac{2am}{2(1+m)d}}\,,~~\beta = \sqrt{\frac{a}{2(1+m)}}\,.
\eea

In the limit $m = 1$ the periodic kink  solution (\ref{2.11}) goes over to the
celebrated kink solution
\be\label{2.14}
\phi(x) = A+B \tanh(\beta x)\,,
\ee
provided
\be\label{2.15}
b = 3 dA\,,~~a = 2 dA^2\,,~~ d B^2 = 2 \beta^2\,,
~~d A^2 = 2\beta^2\,,
\ee
which implies that this solution exists only if
\bea\label{2.16}
&&a, b, d > 0\,,~~2b^2 = 9 a d\,,~~A = \sqrt{\frac{a}{2d}}\,,
\nonumber \\
&&B = \sqrt{\frac{am}{(1+m)d}}\,,~~\beta = \frac{\sqrt{a}}{2}\,.
\eea
Notice that unlike the three pulse solutions, the two kink solutions are valid
if $a, b, d > 0$.

We now show that the asymmetric $\phi^4$ field Eq. (\ref{2.1}) also admits four 
novel  periodic solutions which can be written as superposition of either the
periodic kink or the periodic pulse solution $\sn(x,m)$ or $\dn(x,m)$,  
respectively.

{\bf Solution I}

It is straightforward to check that the asymmetric $\phi^4$ field 
Eq. (\ref{2.1}) admits the periodic solution
\be\label{2.17}
\phi(x) = D -\frac{A \dn(\beta x,m) \cn(\beta x,m)}{1+B\cn^2(\beta x,m)}\,,
~~B > 0\,,
\ee
provided
\be\label{2.18}
0 < m < 1\,,~~b = 3 d D\,,~~a = 2 d D^2\,,~~2b^2 = 9ad
\ee 
and essentially following the arguments used in obtaining a similar solution
for the symmetric $\phi^4$ field Eq. (\ref{1.1}), we find that 
\be\label{2.19}
B = \frac{\sqrt{m}}{1-\sqrt{m}}\,,~~d D^2 = [1+m+6\sqrt{m}]\beta^2 > 0\,,
~~d A^2 = \frac{8 \sqrt{m} \beta^2}{(1-\sqrt{m})^2}\,.
\ee
Notice that for this solution $a, b, d > 0$ 
and it is only valid for $0 < m < 1$, i.e. there
is no corresponding hyperbolic pulse solution. 

On using the identity (\ref{1.19}) one can
re-express the periodic solution I given by (\ref{2.17}) 
as the superposition of two periodic kink solutions, i.e.
\be\label{2.20}
\phi(x) = \sqrt{\frac{a}{2d}} -\frac{\sqrt{2m} \beta}{\sqrt{d}} 
\bigg [\sn(\beta x +\Delta, m) -\sn(\beta x- \Delta, m) \bigg ]\,,
\ee
Here $\Delta$ is defined by $\sn(\sqrt{m}\Delta,1/m) = \pm m^{1/4}$,
where use has been made of the identity (\ref{1.21}). It is worth noting that 
for the periodic kink solution (\ref{2.11}) as well as for the superposed 
periodic kink solution (\ref{2.17}), $a, b, d > 0$.

{\bf Solution II}

Remarkably, the asymmetric $\phi^4$ Eq. (\ref{2.1}) also admits 
another periodic solution
\be\label{2.21}
\phi(x) = D -\frac{A \sn(\beta x,m)}{1+B\cn^2(\beta x,m)}\,,~~B > 0\,.
\ee
On essentially following the arguments used in obtaining a similar solution
for the symmetric $\phi^4$ field Eq. (\ref{1.1}), we find that (\ref{2.21})
is an exact solution provided Eq. (\ref{2.18}) is satisfied and further if 
\be\label{2.22}
B = \frac{\sqrt{m}}{1-\sqrt{m}}\,, ~~~d D^2 = [(1+m) -6\sqrt{m}]\beta^2\,,
~~d A^2 = -8\sqrt{m} \beta^2\,.
\ee
Thus, for this solution $a, b, d < 0$ and only those values
of $m$ are allowed satisfying $6\sqrt{m} > 1+m$. 
Further it is only valid for $0 < m < 1$, i.e. there
is no corresponding hyperbolic pulse solution. 

On using the identity (\ref{1.28}), 
the periodic solution II given by Eq. (\ref{2.21}) can be
re-expressed as the superposition of two periodic kink solutions
\be\label{2.23}
\phi(x) = \sqrt{\frac{|a|}{2|d|}} - i\sqrt{\frac{2m}{|d|}}\beta  
\bigg [\sn(\beta x +\Delta, m)+\sn(\beta x -\Delta, m) \bigg ]\,,
\ee
Here $\Delta$ is defined by $\sn(\sqrt{m}\Delta,1/m) = \pm m^{1/4}$,
where use has been made of the identity (\ref{1.21}).

It is worth noting that while $a, b, d > 0$ for the superposed solution I, 
for the superposed solution II, the values of $a, b, d < 0$. However, the 
value of $B$ is the same for both the solutions.

{\bf Solution III}

It is easy to check that the asymmetric $\phi^4$ field Eq. (\ref{2.1}) 
also admits another periodic solution
\be\label{2.24}
\phi(x) = D -\frac{A \sn(\beta x,m) \cn(\beta x,m)}{1+B\cn^2(\beta x,m)}\,.
\ee
On essentially following the arguments used in obtaining a similar solution
for the symmetric $\phi^4$ field Eq. (\ref{1.1}), we find that (\ref{2.24})
is an exact solution provided Eq. (\ref{2.18}) is satisfied and further if
\bea\label{2.25}
&&B = \frac{1-\sqrt{1-m}}{\sqrt{1-m}}\,,~~
d D^2  = -[(2-m) -6\sqrt{1-m}]\beta^2\,, \nonumber \\
&&d A^2 = -\frac{8(1-\sqrt{1-m})^2 \beta^2}{\sqrt{1-m}}\,.
\eea
Notice that for this solution $a, b, d < 0$ and only those values of $m$ 
are allowed such that  $2-m > 6\sqrt{1-m}$. 
Further, it is only valid for $0 < m < 1$, i.e. there
is no corresponding hyperbolic pulse solution. 

On using the identity (\ref{1.37}), the solution 
III of the asymmetric $\phi^4$ field Eq. (\ref{2.1}) given by Eq. (\ref{2.24}) 
can be re-expressed as a 
superposition of the two periodic pulse solutions, i.e.
\be\label{2.26}
\phi(x) = \sqrt{\frac{|a|}{2|d|}} -\beta \sqrt{\frac{2}{|d|}} 
\bigg (\dn[\beta x -\frac{K(m)}{2},m] -\dn[\beta x +\frac{K(m)}{2},m] \bigg )\,.
\ee

{\bf Solution IV}

The asymmetric $\phi^4$ Eq. (\ref{2.1}) also admits another 
periodic solution
\be\label{2.27}
\phi(x) = D- \frac{A \dn(\beta x,m)}{1+B\cn^2(\beta x,m)}\,.
\ee
On essentially following the arguments used in obtaining a similar solution
for the symmetric $\phi^4$ field Eq. (\ref{1.1}), we find that (\ref{2.27})
is an exact solution provided Eq. (\ref{2.18}) is satisfied and further if
\be\label{2.28}
B = \frac{1-\sqrt{1-m}}{\sqrt{1-m}}\,,~~d D^2 = -[2-m+6\sqrt{1-m}]\beta^2\,,
~~d A^2 = -\frac{8}{\sqrt{1-m}} \beta^2\,.
\ee
Thus for this solution $a, b, d < 0$.
Further, it is only valid for $0 < m < 1$, i.e. there
is no corresponding hyperbolic pulse solution. 

On using the identity (\ref{1.41}), the periodic solution IV 
given by Eq. (\ref{2.27}) can be
re-expressed as the superposition of two periodic pulse solutions, i.e.
\be\label{2.29}
\phi(x) = \sqrt{\frac{|a|}{2|d|}} -\sqrt{\frac{2}{|d|}} \beta  
\bigg (\dn[\beta x +\frac{K(m)}{2}, m]+\dn[\beta x-\frac{K(m)}{2}, m] \bigg )\,.
\ee

It is worth noting that for both the superposed periodic pulse 
solutions III and IV, not only the 
value of $B$ is the same but also $a, b, d < 0$ for both the solutions. 
It is worth reminding that even for the periodic pulse solution as given by
Eqs. (\ref{2.5}) and (\ref{2.6}), $a, b, d < 0$.

\subsection{Novel Superposed (Hyperbolic) Kink Solutions}

We now show that unlike the symmetric $\phi^4$ field theory characterized 
by Eq. (\ref{1.1}), 
the  asymmetric field theory characterized by the field Eq. (\ref{2.1})
admits the superposition of two $\tanh$-type localized kink solutions. 

In particular, we now show that Eq. (\ref{2.1}) admits two distinct 
pulse solutions
\be\label{2.30}
\phi(x) = D - \frac{A}{B+\cosh^2(\beta x)}\,,~~B > 0\,,
\ee
i.e. one with $D \ne 0$ and other with $D = 0$.

{\bf Case I: $D \ne 0$}

On using the ansatz (\ref{2.30}) in Eq. (\ref{2.1}) we find that it is an 
exact solution provided 
\bea\label{2.31}
&&a-b D + d D^2 = 0\,,~~d A^2 = 8B(B+1)\beta^2\,,~~4\beta^2 = d D^2 -a\,,
\nonumber \\
&&(3dD-b)A = 6(2B+1)\beta^2\,.
\eea
Note that for this solution $a, b, d > 0$. 

Now starting from the novel identity (\ref{1.19}) and taking $m = 1$, we 
obtain the corresponding trigonometric identity
\be\label{2.34}
\tanh(y+\Delta) -\tanh(y-\Delta) = \frac{\sinh(2\Delta)}
{B+\cosh^2(y)}\,,~~B = \sinh^2(\Delta)\,.
\ee
On using Eqs. (\ref{2.31}) and (\ref{2.34}) in Eq. (\ref{2.30}) one can 
re-express the solution (\ref{2.30}) 
as a localized state with nonzero asymptote at $\pm \infty$
\be\label{2.35}
\phi(x) = \frac{b+\sqrt{b^2-4ad}}{2d} - \sqrt{\frac{2}{d}} \beta
\bigg [\tanh(\beta x+\Delta) -\tanh(\beta x-\Delta) \bigg ]\,,
\ee
where $\sinh(\Delta) = \sqrt{B}$\,. It is worth noting that while the kink
solution (\ref{2.14}) exists when $b^2 = 9 ad$, the above superposition of 
two kink solutions exists only when $b^2 > 9ad$.

{\bf Case II: $D = 0$}

We now show that Eq. (\ref{2.1}) also admits another pulse solution
\be\label{2.36}
\phi(x) = \frac{A}{B+\cosh^2(\beta x)}\,,~~B > 0\,,
\ee
provided
\be\label{2.37}
\beta = \frac{\sqrt{a}}{2}\,,~~d A^2 =8B(B+1)\beta^2\,,
~~\frac{4(B+1)}{(B+2)^2} = \frac{9ad}{2b^2} < 1\,.
\ee
On using the identity (\ref{2.34}), one can re-express the pulse
solution (\ref{2.36}) as a localized state with zero asymptote at $\pm \infty$
\be\label{2.38}
\phi = \sqrt{\frac{2}{d}} \beta \bigg [\tanh(\beta x +\Delta) 
-\tanh(\beta x - \Delta) \bigg ]\,,
\ee
where $\sinh(\Delta) = \sqrt{B}$\,.
Note that a la kink solution (\ref{2.14}), 
for the solutions (\ref{2.30}) and (\ref{2.36}) too $a, b, d > 0$ 
but whereas the single kink
solution exists only when $2b^2 = 9a d$ while the above superposed
solutions (\ref{2.30}) and (\ref{2.36}) exist only when $2b^2 > 9ad$.

\subsection{Novel Solutions of $\phi^{2}$-$\phi^{2m+2}$-$\phi^{4m+2}$ Field Theory}

Remarkably, it turns out that some of the results of the 
$\phi^{2}$-$\phi^{3}$-$\phi^{4}$  field 
theory are immediately extended to the $\phi^2$-$\phi^{2m+2}$-$\phi^{4m+2}$ 
field theory. In particular, we now show that 
\be\label{2.40}
\phi_{xx} = a\phi - b \phi^{2m+1} + d\phi^{4m+1}\,,~~a, b, d > 0\,,
\ee
also admits the superposition of two $\tanh$-type localized solutions. 
Here $m$ can take any integer or half-integer value. Notice that for 
$m = 1/2$, Eq. (\ref{2.40}) goes over to Eq. (\ref{2.1}) 
for the $\phi^{2}$-$\phi^{3}$-$\phi^{4}$
field theory. Before deriving such a solution, let us first note that in case 
$(2m+1)b^2 = 4(m+1)^2 ad$, Eq. (\ref{2.40}) admits the kink solution 
\cite{csk}
\be\label{2.41}
\phi = \bigg (\frac{(2m+1)a}{4d} \bigg )^{1/4m} [1+\tanh(ax)]^{1/2m}\,.
\ee
As expected, at $m = 1/2$ this kink solution goes over to the kink solution
(\ref{2.14}) for the $\phi^{2,3,4}$ field theory. 
Note that for the kink solution (\ref{2.14}), $9ad = 2b^2$. 

Now we show that in case $4ad(m+1)^2 < (2m+1) b^2$, 
the field Eq. (\ref{2.40}) admits a $\tanh$-type localized superposed 
solution. In particular, it is easy to show that Eq. (\ref{2.40})
also admits the pulse solution
\be\label{2.42}
\phi(x) = \frac{A}{[B+\cosh^2(\beta x)]^{1/2m}}\,,~~B > 0\,,
\ee
provided
\be\label{2.43}
\beta = m \sqrt{a}\,,~~A = \big [\frac{(2m+1)B(B+1)\beta^2}
{d m^2} \big ]^{1/4m}\,,
~~\frac{4(B+1)}{(B+2)^2} = \frac{(m+1)^2 ad}{(2m+1) b^2} < 1\,.
\ee
On using the identity (\ref{2.34}), one can re-express the pulse
solution (\ref{2.42}) as a localized state with zero asymptote at $\pm \infty$
\be\label{2.44}
\phi = \big [\frac{(2m+1)}{4d m^2} \big]^{1/4m} \beta^{1/2m} 
\bigg [\tanh(\beta x +\Delta) -\tanh(\beta x - \Delta) \bigg ]^{1/2m}\,,
\ee
where $\sinh(\Delta) = \sqrt{B}$. 
For $m = 1/2$, all the expressions reduce to those of the 
$\phi^{2}$-$\phi^{3}$-$\phi^{4}$ field 
theory. On the other hand, for $m = 1$ we have the expressions for the 
celebrated $\phi^{2}$-$\phi^{4}$-$\phi^{6}$ field theory. 

It is worth pointing out that for $m= (2p+1)/2, p = 0, 1, 2...$ 
we get theories like $\phi^{2}$-$\phi^{2p+3}$-$\phi^{4p+4}$ field theories, (i.e
for m = 3/2 we get $\phi^2$-$\phi^5$-$\phi^8$ field theories) which are 
not invariant under $\phi \rightarrow -\phi$. 

\section{Novel Superposed Solutions of MKdV Equation}

We now show that the MKdV equation also admits superposed periodic kink and
pulse solutions in addition to the superposed hyperbolic kink 
solution obtained in \cite{tan}. 
Before we do that it is worthwhile reminding about the well known periodic
kink and pulse solutions of the MKdV equation \cite{dj}
\be\label{3.1}
u_t + u_{xxx} + 6 g u^2 u_{x} = 0\,,
\ee 
where $g > (<)$ 0 corresponds to the attractive (repulsive) MKdV equation. 
Without loss
of generality, we will choose $g = 1$ ($-1$) for the attractive (repulsive) MKdV. 

It is well known that Eq. (\ref{3.1}) admits the pulse-like 
solution
\be\label{3.2}
u(x,t) = A \cn(\xi,m)\,,~~\xi = \beta(x-vt)\,,
\ee
provided $g = 1$, i.e. attractive MKdV and further 
\be\label{3.3}
 A^2 = m \beta^2\,,~~v = (2m-1)\beta^2\,.
\ee
In addition, it also admits another periodic pulse solution
\be\label{3.4}
u(x,t) = A \dn(\xi,m)\,,
\ee
provided $g = 1$ and 
\be\label{3.5}
A^2 = \beta^2\,,~~v = (2-m)\beta^2\,.
\ee
In the limit $m = 1$, both these periodic pulse solutions go over to the
(hyperbolic) pulse solution
\be\label{3.6}
u(x,t) = A \sech(\xi)\,,
\ee
provided $g = 1$ and 
\be\label{3.7}
A^2 = \beta^2\,,~~v = \beta^2\,.
\ee

The MKdV Eq. (\ref{3.1}) also admits a periodic kink solution
\be\label{3.8}
u(x,t) = A \sn(\xi,m)\,,
\ee
provided $g = -1$ i.e. repulsive MKdV and 
\be\label{3.9}
A^2 = m \beta^2\,,~~v =  -(1+m) \beta^2\,.
\ee
In the limit $m = 1$, the periodic kink solution goes over to the
(hyperbolic) kink solution
\be\label{3.10}
u(x,t) = A \tanh(\xi)\,,
\ee
provided $g = -1$ and 
\be\label{3.11}
A^2 = \beta^2\,,~~v =  -2 \beta^2\,.
\ee

Let us now show that like the symmetric $\phi^4$ case, the MKdV 
Eq. (\ref{3.1}) also
admits four superposed periodic kink and pulse solutions.
Further, as shown already \cite{tan}, unlike the symmetric $\phi^4$ case, 
the MKdV Eq. (\ref{3.1}) also admits a superposed (hyperbolic) kink solution.

On using the ansatz $u(x,t) = u(\xi)$ where $\xi =\beta(x-vt)$ in the 
MKdV Eq. (\ref{3.1}) and then integrating once, we obtain
\be\label{3.12}
\beta^2 u_{\xi \xi} = v u(\xi) -2g u^3(\xi) + C\,,
\ee
where $C$ is the constant of integration. Now for those solutions for which
$u(\xi)$ as well as $u_{\xi \xi}$ vanish at some value of $\xi = \xi_0$, 
clearly $C = 0$. Thus in such cases Eq. (\ref{3.12}) is essentially similar 
to the symmetric 
$\phi^4$ field Eq. (\ref{1.1}) provided we identify $v$ with $a$, $-2g$ with $d$
and $\xi$ with $\beta x$. In particular $d > (<)$ 0 will correspond in our
case to the repulsive (attractive) MKdV case. It turns out that all the four 
superposed solutions of the symmetric $\phi^4$ 
case are also the solutions of the MKdV Eq. (\ref{3.1}). We list these 
superposed solutions below.

{\bf Solution I}

It is easy to check that the repulsive MKdV Eq. (\ref{3.1}) (i.e. $g = -1$)  
admits the periodic pulse solution
\be\label{3.13}
u(x,t) = \frac{A \dn(\xi,m) \cn(\xi,m)}{1+B\cn^2(\xi,m)}\,,~~B > 0\,,
~~\xi = \beta(x-vt)\,,
\ee
provided
\bea\label{3.14}
&&0 < m < 1\,,~~B = \frac{\sqrt{m}}{1-\sqrt{m}}\,,~~
v = -[1+m+6\sqrt{m}]\beta^2 < 0\,, \nonumber \\
&&A^2 = \frac{4 \sqrt{m} \beta^2}{(1-\sqrt{m})^2}\,.
\eea 
Note that this solution is not valid for $m = 1$, i.e. the MKdV 
Eq. (\ref{3.1}) does not admit a corresponding hyperbolic solution.
Notice that for this solution $v < 0$. 

On using the identity (\ref{1.19}) one can then rewrite the periodic 
pulse solution (\ref{3.13}) as the superposition of the two periodic kink 
solutions, i.e. 
\be\label{3.15}
u(x,t) = \sqrt{2m} \beta 
\bigg [\sn(\xi +\Delta, m) -\sn(\xi- \Delta, m) \bigg ]\,,~~\xi = \beta(x-vt)\,.
\ee
Here $\Delta$ is defined by $\sn(\sqrt{m}\Delta,1/m) = \pm m^{1/4}$,
where use has been made of the identity (\ref{1.21}).

{\bf Remark} 

At this stage we recall the celebrated Miura transformation \cite{miura} which 
showed a remarkable connection between the solutions of the repulsive MKdV 
equation and the KdV equation. In particular, it was shown that if $u$
is a solution of the repulsive MKdV equation
\be\label{3.15a}
u_t +u_{xxx} - 6 u^2 u_x = 0\,,
\ee
then $w = u^2 +u_{x}$ is the corresponding solution of the KdV equation
\be\label{3.15b}
w_t +w_{xxx} - 6 w w_x = 0\,.
\ee
Hence on using the superposed solution (\ref{3.15}) of the repulsive MKdV
equation we obtain the following superposed solution of the KdV Equation
\bea\label{3.15c}
&&w(x,t) = 2m \beta^2 [\sn(\xi +\Delta, m) -\sn(\xi- \Delta, m)]^2 
+ \sqrt{2m} \beta^2 \nonumber \\
&&\big [\cn(\xi+\Delta,m) \dn(\xi+\Delta,m) -
\cn(\xi-\Delta,m) \dn(\xi-\Delta,m) \big ]\,, 
\eea
where $\xi = \beta(x-vt)$. 

A comment is in order here. Unlike the superposed solution (\ref{3.15}), 
Smagin, Tankeyev and Borich \cite{sma} obtained a periodic solution of the
repulsive MKdV Eq. (\ref{3.15a}) which is a product of two
periodic kink solutions, i.e.
\be\label{3.15d}
u(x,t) = D + \sn(\xi+\delta,m) \sn(\xi-\delta,m)\,.
\ee
The nice point of this solution is that in the limit $m=1$ this solution
smoothly goes over to the superposition of two hyperbolic kink solutions
obtained earlier \cite{tan} (which we have mentioned below for completeness).

{\bf Solution II}

The attractive MKdV Eq. (\ref{3.1}) (i.e. $g = 1$) admits the periodic kink 
solution
\be\label{3.16}
u(x,t) = \frac{A \sn(\xi,m)}{1+B\cn^2(\xi,m)}\,,~~B > 0\,,~~
\xi = \beta(x-vt)\,,
\ee
provided
\be\label{3.17}
0 < m < 1\,,~~B = \frac{\sqrt{m}}{1-\sqrt{m}}\,,~~
v = [6\sqrt{m}-(1+m)]\beta^2\,,~~ A^2 = 4\sqrt{m} \beta^2\,.
\ee 
Note that this solution does not exist for $m = 1$, i.e. the corresponding
hyperbolic solution does not exist.

Now on using the novel identity (\ref{1.28}) we can rewrite the solution 
(\ref{3.16}) as the superposition of the two periodic kink solutions, i.e.
\be\label{3.18}
u(x,t) = i \sqrt{m} \beta  
\bigg [\sn(\xi +\Delta, m)+\sn(\xi -\Delta, m) \bigg ]\,,~~\xi = \beta(x-vt)\,. 
\ee
Here $\Delta$ is defined by $\sn(\sqrt{m}\Delta,1/m) = \pm m^{1/4}$,
where use has been made of the identity (\ref{1.21}).

It is worth noting that for both the solutions I and II, the value of $B$ is 
the same but while the solution I is valid in the case of the repulsive MKdV, 
the solution II is valid in the case of the attractive MKdV.

{\bf Solution III}

Yet another periodic solution to the attractive MKdV Eq. (\ref{3.1}) is
\be\label{3.19}
u(x,t) = \frac{A \sn(\xi,m) \cn(\xi,m)}{1+B\cn^2(\xi,m)}\,,~~B > 0\,,~~
\xi = \beta(x-vt)\,,
\ee
provided
\bea\label{3.20}
&&0 < m < 1\,,~~B = \frac{1-\sqrt{1-m}}{\sqrt{1-m}}\,,~~
v = (2-m-6\sqrt{1-m})\beta^2\,, \nonumber \\
&&A^2 = \frac{4(1-\sqrt{1-m})^2 \beta^2}{\sqrt{1-m}}\,.
\eea
On using the identity (\ref{1.37}), the periodic solution (\ref{3.19})  
can be re-expressed as a 
superposition of the two periodic pulse solutions, i.e.
\be\label{3.21}
u(x,t) = \beta \bigg (\dn[\xi -\frac{K(m)}{2},m] 
- \dn[\xi +\frac{K(m)}{2},m] \bigg )\,,~~\xi = \beta(x-vt)\,.
\ee

{\bf Solution IV}

Another periodic solution to the attractive MKdV Eq. (\ref{3.1}) is 
\be\label{9.1}
u(x,t) = \frac{A \dn(\xi,m)}{1+B\cn^2(\xi,m)}\,,~~B > 0\,,~~
\xi = \beta(x-vt)\,,
\ee
provided
\bea\label{9.2}
&&0 < m < 1\,,~~ B = \frac{1-\sqrt{1-m}}{\sqrt{1-m}}\,,
\nonumber \\
&&v = [2-m+6\sqrt{1-m}]\beta^2\,,~~A^2 = \frac{4}{\sqrt{1-m}} \beta^2\,.
\eea
Thus for this solution $v > 0$.

On using the identity (\ref{1.45}) 
the periodic solution IV given by Eq. (\ref{9.1}) can be
re-expressed as superposition of two periodic pulse solutions, i.e.
\be\label{9.3}
u(x,t) = \beta  
\bigg [\dn(\xi +\frac{K(m)}{2}, m) +\dn(\xi -\frac{K(m)}{2}, m) \bigg ]\,,
~~\xi = \beta(x-vt)\,.
\ee

\subsection{Superposed hyperbolic Kink Solution}

For completeness, we now give the hyperbolic pulse solution of the 
repulsive MKdV Eq. (\ref{3.1}) which, as has been shown in \cite{tan},
can be re-expressed as the superposition of the two hyperbolic kink 
solutions. In particular
it is easy to verify that another solution to Eq. (\ref{3.1}) is
\be\label{3.22}
u(x,t) = 1-\frac{A}{B+\cosh^2(\xi)}\,,~~B > 0\,,~~\xi = \beta(x-vt)\,,
\ee
provided $g = -1$, i.e. repulsive MKdV and further 
\be\label{3.23}
A = 2\sqrt{B(B+1)}\beta\,,~~\beta^2 = \frac{4(B+1)}{(B+2)^2} < 1\,,~~
v = 4\beta^2 -6\,.
\ee
On using the identity (\ref{2.34}), the solution (\ref{3.22}) can be
written as the superposition of two (hyperbolic) kink solutions
\be\label{3.25}
u(x,t) = 1-\beta \bigg [\tanh(\xi+\Delta) -\tanh(\xi-\Delta) \bigg ]\,,~~
\xi = \beta(x-vt)\,,
\ee
where $\sinh(\Delta) = \sqrt{B}$.

{\bf Remark}

On using the superposed solution (\ref{3.25}) of the repulsive MKdV Eq. (\ref{3.15a}) 
and using the Miura transformation \cite{miura}, we immediately 
obtain the corresponding superposed solution of the KdV Eq. (\ref{3.15b})
\bea\label{3.25a}
&&w(x,t) = \bigg (1-\beta[\tanh(\xi+\Delta) -\tanh(\xi-\Delta)] \bigg )^2 
\nonumber \\
&&-\beta^2 \bigg [\sech^2(\xi+\Delta) -\sech^2(\xi-\Delta) \bigg ]\,.
\eea

\section{Superposed Solutions of Mixed KdV-MKdV Equation}

We now show that even the mixed KdV-MKdV equation \cite{mixed} 
\be\label{4.1}
u_t + u_{xxx} + 2b u u_{x} + 6 g u^2 u_{x} = 0\,,
\ee
admits superposed periodic kink and pulse solutions. 
But before that let us note the well known 
periodic kink and pulse solutions of the mixed KdV-MKdV 
Eq. (\ref{4.1}). Without loss of 
generality we choose $g = \pm 1$ with $g = 1$ ($-1$) corresponding to the 
attractive (repulsive) MKdV. 

It is well known that Eq. (\ref{4.1}) admits the periodic pulse solution
\be\label{4.2}
u(x,t) = A [1 \pm \cn(\xi,m)]\,,~~\xi = \beta(x-vt)\,,
\ee
in the case of attractive MKdV, i.e. $g = 1$ and further
\be\label{4.3}
A^2 = m \beta^2\,,~~b = -6 \sqrt{m} \beta < 0\,,~~v = -(4m+1)\beta^2\,.
\ee
The attractive MKdV admits another periodic pulse solution
\be\label{4.4}
u(x,t) = A [1 \pm \dn(\xi,m)]\,,
\ee
provided
\be\label{4.5}
A^2 = \beta^2\,,~~b = -6 \beta < 0\,,~~v = -(4+m)\beta^2\,.
\ee
In the limit $m = 1$, both these periodic pulse solutions go over to
the hyperbolic pulse solution
\be\label{4.6}
u(x,t) = A [1 \pm \sech(\xi)]\,,
\ee
provided
\be\label{4.7}
A^2 = \beta^2\,,~~b = -6 \beta < 0\,,~~v = -5\beta^2\,.
\ee

On the other hand, the repulsive MKdV ($g = -1$) admits the periodic 
kink solution
\be\label{4.8}
u(x,t) = A [1 \pm \sn(\xi,m)]\,,
\ee
provided
\be\label{4.9}
A^2 = m \beta^2\,,~~b = 6\sqrt{m} \beta > 0\,,~~v = (5-m)\beta^2\,.
\ee
In the limit $m = 1$, the periodic kink solution (\ref{4.8}) 
goes over to the kink solution
\be\label{4.10}
u(x,t) = A [1 \pm \tanh(\xi)]\,,
\ee
provided
\be\label{4.11}
A^2 =  \beta^2\,,~~b = 6 \beta > 0\,,~~v = 4 \beta^2\,.
\ee

We now show that the mixed KdV-MKdV Eq. (\ref{4.1}) also admits  
periodic superposed kink and pulse solutions which are similar
to those given in Sec. III for the $\phi^{2}$-$\phi^{3}$-$\phi^{4}$ field 
theory and hence we simply list the solutions and omit any details.

On using the ansatz $u(x,t) = u(\xi)$ where $\xi = \beta(x-vt)$ in 
Eq. (\ref{4.1}) and integrating it once we obtain
\be\label{4.11a}
\beta^2 u_{\xi \xi} = v u - b u^{2} -2g u^{3}+C\,,
\ee
where $C$ is the integration constant. 

{\bf Solution I}

It is easy to check that Eq. (\ref{4.1}) with $g = -1$, admits 
the periodic pulse solution
\be\label{4.12}
u(x,t) = D-\frac{A \dn(\xi,m) \cn(\xi,m)}{1+B\cn^2(\xi,m)}\,,~~B > 0\,,
\ee
provided
\be\label{4.12a}
C = -vD + bD^2 +2gD^3\,,
\ee
and further if 
\bea\label{4.13}
&&0 < m < 1\,,~~B = \frac{\sqrt{m}}{1-\sqrt{m}}\,,~~D = \frac{b}{6}\,,~~
\nonumber \\
&&v = \frac{b^2}{6}-[1+m+6\sqrt{m}]\beta^2\,,
~~A^2 = \frac{4 \sqrt{m} \beta^2}{(1-\sqrt{m})^2}\,.
\eea 
Note that this solution is not valid for $m = 1$, i.e. the mixed 
KdV-MKdV Eq. (\ref{4.1}) does not admit a corresponding hyperbolic solution.

On using the identity (\ref{1.19}) one can then rewrite the periodic 
pulse solution (\ref{4.12}) as a superposition of two periodic kink 
solutions, i.e. 
\be\label{4.14}
u(x,t) = \frac{b}{6} -\sqrt{2m} \beta 
\bigg [\sn(\xi +\Delta, m) -\sn(\xi- \Delta, m) \bigg ]\,,~~\xi = \beta(x-vt)\,.
\ee
Here $\Delta$ is defined by $\sn(\sqrt{m}\Delta,1/m) = \pm m^{1/4}$,
where use has been made of the identity (\ref{1.21}).

{\bf Solution II}

It is easy to check that Eq. (\ref{4.1}) with $g = 1$
admits the periodic kink solution
\be\label{4.15}
u(x,t) = D -\frac{A \sn(\xi,m)}{1+B\cn^2(\xi,m)}\,,~~B > 0\,,
\ee
provided $C$ is given by Eq. (\ref{4.12a}) and further if 
\bea\label{4.16}
&&0 < m < 1\,,~~B = \frac{\sqrt{m}}{1-\sqrt{m}}\,,~~D = -\frac{b}{6}\,,
\nonumber \\
&&v = -\frac{b^2}{6} +[6\sqrt{m}-(1+m)]\beta^2\,,
~~ A^2 = 4\sqrt{m} \beta^2\,.
\eea 
Note that this solution does not exist for $m = 1$, i.e. the corresponding
hyperbolic solution does not exist.

Now on using the novel identity (\ref{1.28}) we can rewrite the solution 
(\ref{3.16}) as a superposition of two periodic kink solutions, i.e.
\be\label{4.17}
u(x,t) = -\frac{b}{6} + i \sqrt{m} \beta  
\bigg [\sn(\xi +\Delta, m)+\sn(\xi -\Delta, m) \bigg ]\,,~~\xi = \beta(x-vt)\,.
\ee
Here $\Delta$ is defined by $\sn(\sqrt{m}\Delta,1/m) = \pm m^{1/4}$,
where use has been made of the identity (\ref{1.21}).

It is worth noting that for both the solutions I and II of the mixed KdV-MKdV
Eq. (\ref{4.1}), 
the value of $B$ is the same but while the solution I is valid in case 
$g = -1$, the solution II is valid in case $g = 1$.

{\bf Solution III}

Yet another periodic solution to Eq. (\ref{4.1}) with $g = 1$ is
\be\label{4.18}
u(x,t) = D -\frac{A \sn(\xi,m) \cn(\xi,m)}{1+B\cn^2(\xi,m)}\,,
\ee
provided $C$ is given by Eq. (\ref{4.12a}) and further if 
\bea\label{4.19}
&&0 < m < 1\,,~~B = \frac{1-\sqrt{1-m}}{\sqrt{1-m}}\,,~~D = -\frac{b}{6}\,,
\nonumber \\
&&v = -\frac{b^2}{6} +(2-m-6\sqrt{1-m})\beta^2\,,
~~ A^2 = \frac{4(1-\sqrt{1-m})^2 \beta^2}{\sqrt{1-m}}\,. ~~
\eea

On using the identity (\ref{1.37}), the periodic solution (\ref{4.18})  
can be re-expressed as a 
superposition of two periodic pulse solutions, i.e.
\be\label{4.20}
u(x,t) = -\frac{b}{6} -\beta \bigg (\dn[\xi -\frac{K(m)}{2},m] 
- \dn[\xi +\frac{K(m)}{2},m] \bigg )\,.
\ee

{\bf Solution IV}

Yet another periodic pulse solution to the mixed KdV-MKdV 
Eq. (\ref{4.1}) with $g= 1$ is
\be\label{4.21}
u(x,t) = D- \frac{A \dn(\xi,m)}{1+B\cn^2(\xi,m)}\,,
\ee
provided $C$ is given by Eq. (\ref{4.12a}) and further if
\bea\label{4.22}
&&0 < m < 1\,,~~B = \frac{1-\sqrt{1-m}}{\sqrt{1-m}}\,,
~~D = -\frac{b}{6}\,, \nonumber \\
&&v= -\frac{b^2}{6} +[2-m+6\sqrt{1-m}]\beta^2\,,
~~A^2 = \frac{4}{\sqrt{1-m}} \beta^2\,.
\eea

On using the identity (\ref{1.41}), the periodic solution IV 
given by Eq. (\ref{4.21}) can be
re-expressed as the superposition of two periodic pulse solutions, i.e.
\be\label{4.23}
u(x,t) = -\frac{b}{6} -\beta  
\bigg [\dn(\xi +\frac{K(m)}{2}, m)+\dn(\xi -\frac{K(m)}{2}, m) \bigg ]\,,
~~\xi = \beta(x-vt)\,.
\ee

\subsection{Novel Superposed (Hyperbolic) Kink Solutions}

We now show that the mixed KdV-MKdV Eq. (\ref{4.1})
admits two distinct solutions both of which can be re-expressed as a 
superposition of two $\tanh$-type localized kink solutions. We discuss 
these one by one.

{\bf Solution I}

It is easy to check that the mixed KdV-MKdV
Eq. (\ref{4.1}) admits the solution
\be\label{4.24}
u(x,t) = 1-  \frac{A}{B+\cosh^2(\xi)}\,,~~B > 0\,,~~\xi = \beta(x-vt)\,,
\ee
provided $g = -1, C = -v +b -2$, and further if 
\bea\label{4.25}
&&v = 4\beta^2 - 6 +2b\,,~~A^2 = 4B(B+1)\beta^2\,,~~b > 0\,,
\nonumber \\
&&\beta^2 = \frac{4(6-b)^2 B(B+1)}{36(2B+1)^2} < \frac{b^2}{36}\,.
\eea

On using the identity (\ref{2.34}) one can re-express the solution 
(\ref{4.24}) as a localized state with nonzero asymptote at $\pm \infty$
\be\label{4.26}
u(x,t) = 1- \beta
\bigg [\tanh(\xi +\Delta) -\tanh(\xi -\Delta) \bigg ]\,,~~\xi = \beta(x-vt)\,,
\ee
where $\sinh(\Delta) = \sqrt{B}$\,. It is worth noting that while the kink
solution  (\ref{4.10}) exists if $\beta^2 = \frac{b^2}{36}$, 
the above superposition of the two kink 
solutions, exists only when $\beta^2 <\frac{b^2}{36}$.
\vskip 1.1cm 

{\bf Solution II}

We now show that the mixed KdV-MKdV Eq. (\ref{4.1}) also admits another pulse 
solution
\be\label{4.27}
u(x,t) = \frac{A}{B+\cosh^2(\xi)}\,,~~B > 0\,,~~\xi = \beta(x-vt)\,,
\ee
provided $g = -1, C = 0$ and further if 
\be\label{4.28}
v = 4\beta^2\,,~~A^2 = 4B(B+1)\beta^2\,,
~~\beta^2 = \frac{b^2}{36} \frac{4(B+1)}{(2B+1)^2} < \frac{b^2}{36}\,.
\ee
On using the identity (\ref{2.34}), one can re-express the pulse
solution (\ref{4.27}) as a localized state with zero asymptote at $\pm \infty$
\be\label{4.29}
u(x,t) = \beta \bigg [\tanh(\xi +\Delta) -\tanh(\xi - \Delta) \bigg ]\,,~~
\xi = \beta(x-vt)\,,
\ee
where $\sinh(\Delta) = \sqrt{B}$.
Note that whereas for the kink solution (\ref{4.10}), 
$\beta^2 = \frac{b^2}{36}$, the superposed
solution (\ref{4.27}) exists only when $\beta^2 < \frac{b^2}{36}$.

\subsection{Novel Solutions of $u^{m} u_x$-$u^{2m} u_x$ Mixed KdV-MKdV Equation}

Remarkably, it turns out that some of the results of the mixed KdV-MKdV 
Eq. (\ref{4.1}) 
are immediately extended to the generalized $u^{m} u_x$-$u^{2m} u_x$ mixed KdV-MKdV equation 
\be\label{4.30}
u_t+u_{xxx} +2b u^{m} u_{x} + 6g u^{2m} u_x = 0\,,~~m = 1, 2, 3,... \,.
\ee
In particular, we show that Eq. (\ref{4.30}) admits a pulse solution which 
can be re-expressed as 
a superposition of two $\tanh$-type localized solutions. Notice that
for $m = 1$, Eq. (\ref{4.30}) goes over to Eq. (\ref{4.1}) for the mixed 
KdV-MKdV case. 

On using the ansatz $u(x,t) = u(\xi)$ where $\xi = \beta(x-vt)$ in 
Eq. (\ref{4.30}) and integrating it once we obtain
\be\label{4.31}
\beta^2 u_{\xi \xi} = v u -\frac{2b}{m+1} u^{m+1} 
-\frac{6g}{2m+1} u^{2m+1}+C\,,
\ee
where $C$ is the integration constant. We now consider two solutions
for both of which $u, u_{\xi \xi} =0$ at $\xi = -\infty$ so that $C = 0$
for these two solutions.

Before deriving the desired pulse solution, let us first note that 
Eq. (\ref{4.30}) also admits a kink solution
\be\label{4.32}
u(x,t) = A[1+\tanh( \xi)]^{1/m}\,,~~\xi = \beta(x-vt)\,,
\ee
provided $g = -1$ and further if
\bea\label{4.33}
&&v = \frac{4\beta^2}{m^2}\,,~~b A^{m} = \frac{(m+1)(m+2)}{m^2} \beta^2\,,
\nonumber \\
&&b^2 = \frac{6(m+1)(m+2)^2}{(2m+1) m^2}\beta^2\,.
\eea
As expected, at $m = 1$ this kink solution goes over to the kink solution
(\ref{4.10}). 

It is easy to show that Eq. (\ref{4.30}) also admits the pulse solution
\be\label{4.34}
u(x,t) = \frac{A}{[B+\cosh^2(\xi)]^{1/m}}\,,~~B > 0\,,~~\xi = \beta(x-vt)\,,
\ee
provided
\bea\label{4.35}
&&v = \frac{4\beta^2}{m^2}\,,~~b A^{m} 
= \left[\frac{(m+1)(m+2)(2B+1)}{m^2}\right] \beta^2\,, \nonumber \\
&&b^2 = \frac{6(m+1)(m+2)^2 (2B+1)^2}{4B(B+1)(2m+1) m^2}  \beta^2\,.
\eea
On using the identity (\ref{2.34}), one can re-express the pulse
solution (\ref{4.34}) as a localized state with zero asymptote at $\pm \infty$
\be\label{4.36}
u(x,t) = \left[\frac{(m+1)(2m+1)\beta^2}{6 m^2}\right]^{1/2m} 
\bigg [\tanh(\xi +\Delta) -\tanh(\xi -\Delta) \bigg ]^{1/m}\,,
\ee
where $\sinh(\Delta) = \sqrt{B}$. 
For $m = 1$, all the expressions reduce to those of the mixed KdV-MKdV case.

\section{Superposed Solutions For the NLS Equation}

We now show that even the celebrated NLS equation
\be\label{5.1}
i u_t + u_{xx} + g|u|^2 u = 0\,,
\ee
admits the superposed periodic kink and the pulse solutions. Here 
$g = 1$ ($-$1) corresponds to the attractive (repulsive) NLS. But before 
we explore these solutions let us note the well known periodic kink 
and pulse solutions of the NLS Eq. (\ref{5.1}) \cite{dj}. 

One of the well known periodic pulse solutions of NLS is
\be\label{5.2}
u(x,t) = A e^{i\omega t} \dn(\beta x,m)\,,
\ee
provided $g = 1$, i.e. attractive NLS and further if
\be\label{5.3}
A^2 = 2\beta^2\,,~~\omega = (2-m) \beta^2\,.
\ee
Another well known periodic pulse solution of the NLS Eq. (\ref{5.1}) is
\be\label{5.4}
u(x,t) = A e^{i\omega t} \cn(\beta x,m)\,,
\ee
provided $g = 1$, i.e. attractive NLS and further if 
\be\label{5.5}
A^2 = 2 m\beta^2\,,~~\omega = (2m-1) \beta^2\,.
\ee
In the limit $m = 1$ both the periodic pulse solutions go over to
the hyperbolic pulse solution
\be\label{5.6}
u(x,t) = A e^{i\omega t} \sech(\beta x)\,,
\ee
provided $g = 1$ and 
\be\label{5.7}
A^2 = 2\beta^2\,,~~\omega = \beta^2\,.
\ee

On the other hand, the repulsive NLS (i.e. $g = -1$) is known to
admit the periodic kink solution
\be\label{5.8}
u(x,t) = A e^{i\omega t} \sn(\beta x,m)\,,
\ee
provided 
\be\label{5.9}
A^2 = 2 m\beta^2\,,~~\omega = -(1+m) \beta^2\,.
\ee
In the limit $m = 1$, the periodic kink solution (\ref{5.8}) goes
over to the (hyperbolic) kink solution
\be\label{5.10}
u(x,t) = A e^{i\omega t} \tanh(\beta x)\,,
\ee
provided 
\be\label{5.11}
A^2 = 2 \beta^2\,,~~\omega = -2 \beta^2\,.
\ee

We now show that the NLS Eq. (\ref{5.1}) admits three periodic
pulse and kink solutions which can be re-expressed as the 
superposition of the periodic kink and the pulse solutions. 
This is easily seen if one notices that the NLS Eq. (\ref{5.1}) can be 
mapped to the symmetric $\phi^4$ Eq. (\ref{1.1}) provided 
we start with the ansatz
\be\label{5.12}
u(x,t) = e^{i\omega t} \phi(x)\,,
\ee
with $\phi(x)$ being real. On substituting the ansatz (\ref{5.12})
in Eq. (\ref{5.1}) we obtain
\be\label{5.13}
\phi_{xx} = \omega \phi - g \phi^3\,,
\ee
which is identical to the symmetric $\phi^4$ field Eq. (\ref{1.1})
provided we identify $\omega$ with $a$ and $d$ with $-g$. Using this
identification we now simply list those three superposed periodic solutions
of NLS Eq. (\ref{5.1}) in which $\phi$ is real.

{\bf Solution I}

It is easy to check that the NLS field Eq. (\ref{5.1})  admits the periodic 
solution
\be\label{5.14}
u(x,t) = e^{i\omega t} \frac{A \dn(\beta x,m) \cn(\beta x,m)}
{1+B\cn^2(\beta x,m)}\,,~~B > 0\,,
\ee
provided $g = -1$, i.e. repulsive NLS and further if
\bea\label{5.15}
&&0 < m < 1\,,~~B = \frac{\sqrt{m}}{1-\sqrt{m}}\,, ~~
\omega = -[1+m+6\sqrt{m}]\beta^2 < 0\,, \nonumber \\
&&A^2 = \frac{8 \sqrt{m} \beta^2}{(1-\sqrt{m})^2}\,.
\eea
On using the identity (\ref{1.19}), one can
re-express the periodic solution I given by (\ref{5.14}) 
as a superposition of the two periodic kink solutions, i.e.
\be\label{5.16}
u(x,t) = e^{i\omega t} \sqrt{2m} \beta
\bigg [\sn(\beta x +\Delta, m) -\sn(\beta x- \Delta, m) \bigg ]\,. 
\ee
Here $\Delta$ is defined by $\sn(\sqrt{m}\Delta,1/m) = \pm m^{1/4}$.

{\bf Solution II}

It is easy to check that the NLS Eq. (\ref{5.1}) admits another periodic 
solution
\be\label{5.20}
u(x,t) = e^{i\omega t} \frac{A \sn(\beta x,m) \cn(\beta x,m)}
{1+B\cn^2(\beta x,m)}\,,
\ee
provided $g = 1$, i.e. attractive NLS and further if 
\bea\label{5.21}
&&0 < m < 1\,,~~B = \frac{1-\sqrt{1-m}}{\sqrt{1-m}}\,,~~
\omega = (2-m-6\sqrt{1-m})\beta^2\,, \nonumber \\
&&A^2 = \frac{8(1-\sqrt{1-m})^2 \beta^2}{\sqrt{1-m}}\,.
\eea
On using the identity (\ref{1.37}) one can re-express the 
solution (\ref{5.20}) as
a superposition of the two periodic pulse solutions, i.e.
\be\label{5.22}
u(x,t) = e^{i\omega t} \sqrt{2} \beta \bigg (\dn[\beta(x) -\frac{K(m)}{2},m] 
- \dn[\beta(x) +\frac{K(m)}{2},m] \bigg )\,.
\ee

{\bf Solution III}

Another periodic solution to the NLS Eq. (\ref{5.1}) is
\be\label{5.23}
u(x,t) = e^{i\omega t} \frac{A \dn(\beta x,m)}{1+B\cn^2(\beta x,m)}\,,
\ee
provided $g = 1$, i.e. attractive NLS and further
\be\label{5.24}
0 < m < 1\,,~~ B = \frac{1-\sqrt{1-m}}{\sqrt{1-m}}\,,~~
\omega = [2-m+6\sqrt{1-m}]\beta^2\,,~~A^2 = \frac{8}{\sqrt{1-m}} \beta^2\,.
\ee
On using the identity (\ref{1.41}), the periodic solution (\ref{5.23})
can be re-expressed as a superposition of the two periodic pulse 
solutions, i.e.
\be\label{5.25}
u(x,t) = e^{i\omega t} \sqrt{2} \beta  
\bigg (\dn[\beta x +\frac{K(m)}{2}, m]+\dn[\beta x-\frac{K(m)}{2}, m] \bigg )\,.
\ee
It is worth noting that for both the superposed periodic pulse 
solutions II and III, not only the 
value of $B$ is the same but both the solutions are valid for the 
attractive NLS. 

\section{Superposed Solutions of the quadratic-cubic NLS}

We now show that even the quadratic-cubic NLS equation \cite{quad} given 
by
\be\label{6.1}
iu_t + u_{xx} +b|u|u + g |u|^2 u = 0\,,
\ee
admits superposed periodic pulse and the hyperbolic kink solutions.

We start with the ansatz
\be\label{6.2}
u(x,t) = e^{i\omega t} \phi(x)\,,
\ee
where we will only consider those $\phi(x)$ which are nonnegative so that
$|\phi| = \phi$. On substituting the ansatz (\ref{6.2}) in Eq. (\ref{6.1})
it is easy to see that $\phi(x)$ satisfies the field equation 
\be\label{6.3}
\phi_{xx} = \omega \phi - b \phi^2 -g \phi^3\,, 
\ee
which is identical to the asymmetric $\phi^4$ field Eq. (\ref{2.1}) 
provided we identify $\omega$ with $a$ and $-g$ with $d$ while $b$ is 
identical in
both the cases. It then follows that all those solutions of the asymmetric
$\phi^4$ Eq. (\ref{2.1}) in which $\phi(x)$ is nonnegative will also be 
the solutions of the quadratic-cubic Eq. (\ref{6.1}). We now simply
mention such solutions one by one.

Let us first discuss the periodic kink and pulse solutions.
It is easy to show that Eq. (\ref{6.1}) admits the periodic pulse solution
\be\label{6.4}
u(x,t) = e^{i\omega t} [A+ B\dn(\beta x,m)]\,,
\ee
provided $g = 1$ and further if
\be\label{6.5}
b = -3A\,,~~\omega = -2 A^2\,,~~ B^2 = 2 \beta^2\,,
~~ A^2 = (2-m)\beta^2\,. 
\ee
In the limit $m = 1$, the periodic pulse solution (\ref{6.4}) 
goes over to the hyperbolic pulse solution
\be\label{6.6}
\phi(x) = A+ B\sech(\beta x)\,,
\ee
provided $g = 1$ and further if
\be\label{6.7}
b = -3A\,,~~\omega = -2 A^2\,,~~ B^2 = 2 \beta^2\,,
~~ A^2 = \beta^2\,.
\ee

The quadratic-cubic NLS Eq. (\ref{6.1}) also admits a periodic kink
solution
\be\label{6.8}
u(x,t) = e^{i\omega t} [A+ B\sn(\beta x,m)]\,,
\ee
provided $g = -1$ and further if
\be\label{6.9}
b = 3A\,,~~\omega = 2 A^2\,,~~ B^2 = 2m \beta^2\,,
~~ A^2 = (1+m) \beta^2\,. 
\ee
In the limit $m = 1$, the periodic kink solution goes over to
the hyperbolic kink solution
\be\label{6.10}
u(x,t) = e^{i\omega t} [A+ B\tanh(\beta x)]\,,
\ee
provided $g = -1$ and further if
\be\label{6.11}
b = 3A\,,~~\omega = 2 A^2\,,~~ B^2 = 2 \beta^2\,,
~~ A^2 =  2\beta^2\,. 
\ee

Let us now discuss the superposed solutions of Eq. (\ref{6.1}).
On comparing with the four periodic superposed solutions for the 
asymmetric $\phi^4$ discussed in Sec. III it is clear that in this case 
only one superposed periodic solution and one
superposed hyperbolic solution are possible which we mention next.

{\bf Solution I: Periodic Superposed Pulse Solution}

In particular, it is easy to show that
\be\label{6.12}
u(x,t) =  e^{i\omega t} \left[D+\frac{A\dn(\beta x,m)}
{1+B\cn^2(\beta x, m)}\right]\,,
\ee
is an exact solution with $A, D, B$ all being positive provided
$g = 1$ and further if
\bea\label{6.13}
&&D^2 = -\omega = 2[2-m+\sqrt{1-m}] \beta^2\,,
~~B = \frac{1-\sqrt{1-m}}{\sqrt{1-m}}\,, \nonumber \\
&&0 < m < 1\,, ~~~ A^2 = \frac{8 \beta^2}{\sqrt{1-m}}\,.
\eea
On using the identity (\ref{1.41}), the periodic solution (\ref{6.12}) 
can be re-expressed as a superposition of the two periodic pulse 
solutions, i.e.
\be\label{6.14}
u(x,t) = e^{i\omega t} \beta \bigg [\dn(\beta x +\frac{K(m)}{2}, m)
+\dn(\beta x -\frac{K(m)}{2}, m) \bigg ]\,.
\ee

{\bf Solution II: Hyperbolic Superposed Solution}

It is easy to show that the quadratic-cubic NLS Eq. (\ref{6.1}) admits 
the hyperbolic solution
\be\label{6.15}
u(x,t) = e^{i\omega t} \frac{A}{B+\cosh^2(\beta x)}\,,
\ee
provided $g =-1$ and further if
\be\label{6.16}
\omega = 4 \beta^2\,,~~A^2 = 8B(B+1)\beta^2\,,
~~b^2 = \frac{18(2B+1)^2}{4B(B+1)}\beta^2\,.
\ee
On using the identity (\ref{2.34}), one can re-express the pulse
solution (\ref{6.15}) as a localized state with zero asymptote at $\pm \infty$
\be\label{6.17}
u(x,t) = e^{i\omega t} \sqrt{2} \beta 
\bigg [\tanh(\beta x +\Delta) -\tanh(\beta x - \Delta) \bigg ]\,,
\ee
where $\sinh(\Delta) = \sqrt{B}$.

\subsection{Superposed Solutions For $|u|^{m} u$-$|u|^{2m} u$ NLS Equation}

We now show that some of the results of the quadratic-cubic NLS equation 
\cite{quad} are easily extended to the generalized NLS equation
\be\label{7.1}
iu_t+u_{xx}+b|u|^{m} u + g |u|^{2m} u = 0\,,~~m = 1, 2, 3,... \,.
\ee
Note that for $m = 1$ this reduces to the quadratic-cubic NLS equation \cite{quad} 
while for $m = 2$ it reduces to the cubic-quintic NLS equation \cite{quintic}. Before 
we discuss the superposed solution, it is worth pointing out that Eq. (\ref{7.1}) 
admits a kink solution
\be\label{7.2}
u(x,t) = e^{i\omega t} A[1 \pm \tanh(\beta x)]^{1/m}\,,
\ee
provided $g = -1$ and further if
\be\label{7.3}
\omega = \frac{4\beta^2}{m^2}\,, ~~~A^{2m} = \frac{(1+m)\beta^2}{m^2}\,,~~
b^2 = \frac{4(m+2)^2}{(1+m)m^2} \beta^2\,.
\ee

Finally, it is easy to show that the generalized NLS Eq. (\ref{7.1}) also
admits a pulse solution
\be\label{7.4}
u(x,t) = e^{i\omega t} \frac{A}{[B+\cosh^2(\beta x)]^{1/m}}\,,
\ee
provided $g = -1$ and further if
\bea\label{7.5}
&&\omega = \frac{4 \beta^2}{m^2}\,,~~
A^{2m} = \frac{4B(B+1)(m+1)}{m^2} \beta^2\,, \nonumber \\
&&b^2 = \frac{4(m+2)^2 \beta^2}{(1+m)m^2} \frac{(2B+1)^2}{4B(B+1)}\,.
\eea
On using the identity (\ref{2.34}), one can re-express the solution (\ref{3.4}) 
as a superposition of the two kink solutions
\be\label{7.6}
u(x,t) = e^{i\omega t} \left[\frac{(m+1)}{m^2}\right]^{1/2m} \beta^{1/m} 
\bigg [\tanh(\beta x +\Delta) -\tanh(\beta x - \Delta) \bigg ]^{1/m}\,,
\ee
where 
$\sinh(\Delta) = \sqrt{B}$.

\section{Conclusion and Open Problems}

In this paper we have shown that a large number of nonlinear equations such 
as the symmetric and asymmetric (double well) $\phi^4$ equations, MKdV 
equation \cite{dj}, mixed KdV-MKdV equation \cite{mixed}, NLS as well as 
quadratic-cubic NLS \cite{quad}, which have applications in several areas of 
physics, admit novel superposed periodic kink and pulse solutions. 
Besides, some of them also admit hyperbolic superposed kink solutions. 
We have also shown that even generalized higher order neutral scalar field
theory models, generalized mixed KdV-MKdV models as well as generalized 
mixed higher order NLS models  admit superposed hyperbolic kink solutions. 
We believe that we have only touched the tip of the iceberg and there are 
many more surprises in store. This paper raises several questions which are 
still not quite understood. We list some of them below.

\begin{enumerate}

\item In this paper, in several different models we have obtained a
number of solutions which can be re-expressed as 
either the sum or the difference of two $\sn(x,m)$ or two $\dn(x,m)$ 
Jacobi elliptic functions. However, we have not been able to obtain similar 
superposed solutions in the Jacobi elliptic $\cn(x,m)$ case. It is not clear as 
to what is the underlying reason. Clearly it would be worthwhile to find 
$\cn(x,m)$ superposed solutions in some nonlinear models.

\item While we have been able to obtain hyperbolic solutions which can be
re-expressed as difference of two $\tanh(x)$ type solutions, so far we have
not been able to obtain hyperbolic solutions which can be 
re-expressed either as sum of two $\tanh(x)$ or sum or difference of two 
$\sech(x)$ type solutions. It is clearly of interest to look for such solutions. 
\item It is not clear what is the physical interpretation of such 
superposed periodic or hyperbolic solutions. Do they correspond to a 
bound state of two kink or two pulse solutions as in some field theory and 
condensed matter contexts \cite{dashen, campbell, saxena, thies}? Or do 
they merely correspond to some excitation of two kink or two pulse solutions? 
It is worthwhile finding the interpretation of
such superposed solutions vis a vis a single kink or pulse solution.

\item It is clearly of interest to discover other nonlinear equations which 
also admit such or even more unusual superposed solutions.  

\end{enumerate}

Hopefully, one can find answers to some of the questions raised above.

\section{Acknowledgment}

AK is grateful to Indian National Science Academy (INSA) for the
award of INSA Senior Scientist position at Savitribai Phule Pune University. 
The work of AS was supported by the U.S. Department of energy.

\end{document}